\documentclass[showpacs,amssymb]{revtex4}

\usepackage{graphicx}
\usepackage{dcolumn}
\usepackage{bm}

\begin{document}

\title{Nonsingular and accelerated expanding universe from effective Yang-Mills theory}
\author{Vitorio A. De Lorenci}
 \email{delorenci@unifei.edu.br}
  \affiliation{Instituto de Ci\^encias Exatas, Universidade Federal de Itajub\'a, 
37500-903 Itajub\'a, M. G., Brazil,}
  \affiliation{PH Department, TH Unit, CERN, 1211 Geneva 23, Switzerland}

\begin{abstract}
The energy-momentum tensor coming from one-parameter effective Yang-Mills theory
is here used to describe the matter-energy content of the homogeneous and isotropic
Friedmann cosmology in its early stages. 
The behavior of all solutions is examined. 
Particularly, it is shown that only solutions corresponding to an open model
do allow the universe to evolve into an accelerated expansion. This result
appears as a possible mechanism for an inflationary phase produced by a vector field.
Further, depending on the value of some parameters characterizing the system, the
resulting models are classified as singular or nonsingular.
\end{abstract}
\pacs{98.80.-k, 98.80.Cq, 98.80.Bp}
\maketitle

%
%
\section{Introduction}
\label{I}
The Friedmann-Lema\^itre cosmological model \cite{friedmann1922,lemaitre1927}, 
described by the Robertson-Walker geometry \cite{robertson1935,walker1936} 
with classical electrodynamics as its source, 
leads to a singularity at a finite time in the past \cite{Kolb}. It
is usually identified as the origin of the universe. In fact, this
singular behavior points out that, around the very beginning, 
the spacetime curvature becomes arbitrarily large, thus being 
beyond the domain of applicability of the model. It is expected that a
quantum theory for gravity would circumvent the appearance of this
curvature singularity by changing the predictions of the general relativity
in the limit of large curvature.

There are many proposals of cosmological solutions 
without a primordial singularity. Mostly they are based
on the fact that a collapsing phase can achieve a minimum,
then evolving into a expanding phase. These nonsingular solutions are usually
called bouncing cosmology.
Such models are based on a variety of distinct mechanisms, 
such as cosmological constant \cite{DeSitter}, 
non-minimal couplings \cite{Gunzig}, nonlinear Lagrangians 
involving quadratic terms in the curvature \cite{Mukhanov,brandenberger1993,moessen1995},
non-equilibrium thermodynamics \cite{Salim} and loop quantum cosmology
\cite{bojowald2001}, among others
\cite{Elbaz,delorenci2002,Breton,klippert,cai2007,hossain2009}.
In general, the singularity theorems \cite{hawking1973} are circumvented by 
the appearance of a high (but nevertheless finite) 
negative pressure in the early phase of the universe. 
An initial singularity can also be avoided by assuming a quantum
creation of a small but finite universe, and hence a beginning of
time \cite{hartle1983,vilenkin1986}. 
For a recent review on bouncing cosmology see \cite{novello2008}.

The issue of inflationary cosmology has also been largely considered in
the literature. For a recent review on theory and observations, see 
\cite{baumann2008} and references therein. 
Data from the Wilkinson Microwave Anisotropy Probe (WMAP) observations  
\cite{spergel2007,hinshaw2008} have imposed several restrictions to the
possible models describing an inflationary cosmology.
Particularly, evidence \cite{yadav2008} for primordial non-Gaussianity
in the temperature anisotropy of the cosmic microwave background radiation
(CMBR) seems to disfavor canonical single-field slow-roll inflation. 
Further, the spatial curvature of the universe is found to be small
(with an spatial curvature parameter $\Omega_K$ of order $10^{-2}$),
but not so small as predicted by the conventional inflationary models,
where $\Omega_K \sim 10^{-5}$. 
%
%
WMAP data seem to favor cosmological models with flat (Euclidean) 
or open (negative curvature) spatial sections.
If the spatial curvature is non null, even being small, the implications
for the time evolution of the universe can really be important, as
described by general relativity.
Recently \cite{bamba2008}, inflationary cosmology and
late-time accelerated expansion of the universe were considered in the context of
non-minimally coupled effective Yang-Mills (Y-M) fields. It was shown that Y-M
fields with a non-minimal gravitational coupling can produce an
accelerated expanding phase in a flat (Euclidean spatial section) universe 
model. A complete analysis of the consequences for cosmology of the minimal 
coupling between gravity and effective Y-M fields has not been considered so far in the literature.
The issue of inflation appears also in the context of  nonsingular quantum
cosmological models \cite{falciano2008}, where the inflationary phase appears
as a quantum cosmological effect, instead of being produced by an inflaton field.  

In this paper, the cosmological consequences of the minimal coupling between gravity and
effective Y-M theory are investigated. The energy-momentum tensor coming from 
the effective Y-M fields is assumed to be the dominant source of the matter-energy content of 
the early universe, which is assumed to be homogeneous and isotropic.
The behavior of all possible solutions is examined for the three distinct 
topologies, corresponding to flat, closed (positive curvature) 
and open (negative curvature) spatial sections. 
It is shown that when Euclidean or closed sections are considered, the solutions
presenting expansion are decelerated and evolve into a static configuration. From
this static phase the system can evolve into a collapsing phase 
if fluctuations on the scale factor are allowed. 
On the other hand, solutions corresponding to an open model 
do allow the universe to evolve into an accelerated expansion, thus providing 
a simple mechanism for an early inflationary phase produced
by a vector field. Models presenting an inflationary phase produced by vector fields
have been called as dark energy models \cite{elizalde2003,zhao2006,xia2007}. 
A graceful exit from the inflationary phase appears naturally because of a relationship
between the scale factor and the magnitude of the Y-M field. 
Further, depending on the value of some parameters that characterize the system, the
solutions turn out to be singular or nonsingular.
Solutions exhibiting collapsing phases are also examined.

The paper is organized as follows: the following section presents some general aspects
of the homogeneous and isotropic cosmology, including the main
equations. In Sec. \ref{III}, the energy-momentum tensor for general one-parameter 
Lagrangian density is derived. It is also presented the procedure of field
averaging which renders the system a perfect fluid configuration. Section \ref{IV}
deals with effective Y-M theory. In this context, the equation 
governing the time evolution of the scale factor is derived. In Sec. \ref{V},  
cosmological models resulting from the minimal coupling between gravity and 
effective Y-M theory are examined. There, a suitable functional form for
the effective coupling is assumed. In Sec. \ref{VI}, we use the results obtained before
in order to discuss the regime of small 
coupling with large mean fields, for which the effective coupling presents 
a well known form. Finally, concluding remarks are presented in Sec. \ref{VII}.

Heaviside non-rationalized units are used.  
Latin indices run in the range $(1,2,3)$ and 
Greek indices run in the range $(0,1,2,3)$. Throughout the text the units 
of $c=1=\hbar$ are used, unless otherwise stated.

%
%
\section{Friedmann-\-Lema\^itre-\-Robertson-\-Walker universe}
\label{II}
Maxwell electrodynamics as source of gravity in the homogeneous and
isotropic Friedmann-Lema\^itre cosmology \cite{friedmann1922,lemaitre1927} 
leads to singular universe models.
In this framework, this is a direct consequence of the 
singularity theorems \cite{hawking1973}, 
and follows from energy conservation law
and Raychaudhuri equation \cite{raychaudhuri1955}. 
Let us assume the homogeneous and isotropic Friedmann-Lema\^itre 
model as described by the Robertson-Walker
metric \cite{robertson1935,walker1936} in the form
\begin{equation}
ds^2 = dt^2 - \frac{A^2(t)}{(1+\epsilon r^2/4)^2}\,
\left[dr^2+r^2\,(d\theta^2+\sin^2\theta\,d\varphi^2)\right],
\label{metric}
\end{equation}
where $\epsilon=-1,\,0,\,+1$ hold for 
the open, flat (or Euclidean) and closed sections, respectively.
Here $(r,\theta,\varphi)$ are dimensionless comoving coordinates. 
In what follows the above mentioned framework will be simply referred as 
FLRW\footnote{Causal geodesic completeness of FLRW spacetime was recently considered
in \cite{fernandez2006}.}. The 3-dimensional surface of homogeneity $t=constant$ 
is orthogonal to a fundamental class of observers 
determined by a four-velocity vector field 
$v^{\mu} = \delta^{\mu}_{0}$.  

For a perfect fluid with energy density $\rho$ and pressure $p$, 
the energy conservation law and the Raychaudhuri equation read,
respectively,
\begin{eqnarray}
&&\dot{\rho} + 3(\rho + p) \frac{\dot{A}}{A} = 0,
\label{27}
\\
&&\frac{\ddot{A}}{A} = - \,\frac{\kappa}{6} (\rho + 3p),
\label{28}
\end{eqnarray}
in which $\kappa$ is the Einstein gravitational constant ($\kappa = 8\pi G$) 
and `dot' denotes Lie derivative respective to $v^\mu$, 
that is $\partial/\partial t$. It is worth to mention here that
$\rho + 3p \ge 0$ states the strong energy-dominance
condition (SEC). For instance, in the case of $A(t)$ describing an
expanding phase, satisfying SEC means that $\ddot{A} < 0$. In this case we
say that gravity decelerates the expansion. On the other hand, 
if this condition is violated the corresponding model will
present a phase of accelerated expansion, as required for an
inflationary cosmology. This issue will be further addressed
when cosmological models are examined.

Multiplying Eq. (\ref{28}) by $A\dot{A}$ we obtain
\begin{equation}
\dot{A}\ddot{A} +  \frac{k}{6} A\dot{A} (\rho + 3p) =0.
\label{28a}
\end{equation}
But $A\dot{A} (\rho + 3p) = -(A^2\rho)^{\mbox{\bf$\cdot$}}$, where Eq. (\ref{27})
was used. Thus, introducing this result in Eq. (\ref{28a}), and after 
integrating it, we obtain the Friedmann equation\footnote{Usually, the 
Friedmann equation is presented in terms
of the density parameter $\Omega \doteq (\kappa \rho)/(3H^2)$ as
$\Omega -1 = \epsilon/A^2 H^2$, were $H$ is the Hubble 
parameter $H\doteq \dot{A}/A$.}
\begin{equation}
\frac{\kappa}{3}\,\rho=\left(\frac{\dot{A}}{A}\right)^2
+\frac{\epsilon}{A^2},
\label{29}
\end{equation}
which consists of a first integral of Eqs.\ (\ref{27}) and (\ref{28}).
If $\rho$ is a known function
of $A(t)$, the above equation will determine the solutions for the
scale factor. In other words, the time evolution of the scale factor is
determined by Einstein's equations in terms of the matter-energy content
of the universe. 
%
%
\section{Energy-momentum tensor for one-parameter nonlinear Lagrangian}
\label{III}
This section provides a description of the tensor characterizing 
the matter-energy contents of the system, which is assumed to be described 
by a general one-parameter Lagrangian density. We begin by introducing
some well known objects, which will be useful to
set the notation. The strength tensor field $F_{\mu\nu}^{(a)}$ and the gauge 
field $A_{\mu}^{(a)}$ are related by 
\begin{equation}
F_{\mu\nu}^{(a)} = \partial_{\mu}A_{\nu}^{(a)} -  \partial_{\nu}A_{\mu}^{(a)} 
+ C^{abc}A_{\mu}^{(b)}A_{\nu}^{(c)},
\label{4a}
\end{equation}
where $C^{abc}$ represents the structure constant for a compact Lie group $G$.
This tensor field can be conveniently defined in terms of the non-Abelian
electric $E_\mu^{(a)}$ and magnetic $B_\mu^{(a)}$ color fields as
\begin{equation}
F_{\mu\nu}^{(a)} = v_{\mu}E_{\nu}^{(a)} -  v_{\nu}E_{\mu}^{(a)} - \eta_{\mu\nu}{}^{\alpha\beta}
v_{\alpha}B_{\beta}^{(a)}.
\label{4}
\end{equation}
In order to alleviate the notation the `color' indices, above indicated by
upper brackets, will be omitted in what follows.

Let us assume the gauge invariant Lagrangian density to be a general function of the
Lorentz invariant $F \doteq F^{\mu\nu}F_{\mu\nu} = 2(B^2-E^2)$ as $L = L(F)$.
The energy-momentum tensor for this class of one-parameter theories can
be presented as 
\begin{equation}
T_{\mu\nu} = -4 L_F F_{\mu\,\alpha}\,F^{\alpha}\mbox{}_{\nu} - L \,g_{\mu\nu}, 
\label{1}
\end{equation}
in which $L_{F}$ represents the derivative of the
Lagrangian density with respect to the invariant $F$.

Since the spatial sections of FLRW geometry are isotropic, 
we may consider that the color fields can generate such universe 
only if an averaging procedure is performed \cite{tolman1930}. 
For this propose the system is assumed to
satisfy the following requirements:
(i) the volumetric spatial average of the color field strength does not depend on directions;
(ii) it is equally probable that the products ${E^i E^j}$,  
${B^i B^j}$  and ${E^i B^j}$  (with $i \neq j$), at any time, 
take positive or negative values;
(iii) there is no net flow of energy as measured by comoving observers. 
The above mentioned volumetric spatial average of an arbitrary quantity $X$ 
for a given instant of time $t$ is defined as  
\begin{equation}
\label{Ref-1}
\left< X \right> \doteq
\lim_{V\rightarrow V_0}
\frac{1}{V}\int X\,\sqrt{-g}\,d^3\!x^i,
\end{equation}
with $V=\int\sqrt{-g}\,d^3\!x^i$, and $V_0$ stands for the time dependent volume of the whole space.  
Similar average procedures have already been considered in
\cite{tolman1930,delorenci2002,kunze2008,delorenci2008}.

In terms of the electric and magnetic color fields, these requirements imply that
\begin{eqnarray}
\left< E_i \right>
= &0&, \qquad 
%
\left< B_i \right>
= 0,\qquad
\label{2b}\\[1ex]
%
\left< E_i\, B_j \right>
&=&  -\, \frac{1}{3} (\vec{E}\cdot\vec{B}) \,g_{ij},
\label{2c}\\[1ex]
%
\left< E_i\, E_j \right>
&=& -\, \frac{1}{3} E^2 \,g_{ij},
\label{2d}\\[1ex]
%
\left< B_i\, B_j \right>
&=&  -\, \frac{1}{3} B^2 \,g_{ij},
\label{2e}
\end{eqnarray}
where we have defined $E^2\doteq -E^i E_i$ and $B^2\doteq -B^i B_i$.
Note that Eq. (\ref{2c}) implies in $\left< E_i\, B_j - B_i\, E_j \right> = 0$.

Applying the above average procedure to the energy-momentum tensor we obtain the
following non null components:
\begin{eqnarray}
\left< T_{00} \right> &=& -(4L_F E^2 + L) g_{00},
\label{9}
\\
\left< T_{ij} \right> &=& -\left[\frac{4}{3}L_F(E^2-2 B^2) + L \right]g_{ij},
\label{10}
\end{eqnarray}
which can be presented as a perfect fluid configuration 
with energy density $\rho$ and pressure $p$ as 
\begin{equation}
\left< T_{\mu\nu} \right> 
= (\rho + p)\, v_{\mu}\, v_{\nu} 
- p\, g_{\mu\nu},
\label{13}
\end{equation}
where we identify:
\begin{eqnarray}
\rho &=& -4L_F E^2 - L,
\label{14}
\\
p &=& \frac{4}{3}(E^2 - 2 B^2)L_F + L.
\label{15}
\end{eqnarray}

For the particular case of Maxwell electrodynamics, with $L=-F/4$, we obtain the classical
result $\rho = 3 p = (E^2+B^2)/2$ (here $E$ and $B$ stand for the Abelian 
electric and magnetic fields). The fact that 
both the energy density and the pressure are positive definite 
for all times yields the singular nature of FLRW universes. 
The Einstein equations for the above energy-momentum configuration 
lead to \cite{robertson1933} the classical solution for the scale factor 
$A(t)= (A_o^2t-\epsilon t^2)^{1/2}$,
where $A_o$ is an arbitrary constant.  

For the case of nonlinear spin-one fields described by the Lagrangian density 
$L=-F/4 + \alpha F^2 + \beta G^2$, which encompass the
first order terms coming from one-loop QED, it can be shown \cite{delorenci2002} that,
in the absence of electric field,
$\rho = (B^2/2)(1-8\alpha B^2)$ and $p = (B^2/6)(1-40\alpha B^2)$. 
This toy model presents solutions exhibiting a nonsingular behavior for the scale factor. 
Regards must be taken in applying such model to the description of the
early universe. Higher order terms coming from the field
expansion in the one-loop effective Lagrangian for QED may not be negligible in the regime 
of large mean fields, as it would be expected to occur in the early universe. Nevertheless,    
it could be considered in a specific phase where the above terms in the Lagrangian
density would correspond to the dominant terms in the effective QED Lagrangian.

Let us return to the general case determined by Eq. (\ref{13}). Taking in
consideration the averaging procedure on fields, 
the energy conservation law stated by Eq. (\ref{27}) reduces to 
\begin{eqnarray}
&&\frac{4L_{FF}}{L_F}E^2\frac{\partial}{\partial t}(B^2-E^2)
\nonumber
\\ 
&&+ (B^2+E^2)\frac{\partial}{\partial t}\ln\left[(B^2+E^2)A^4\right]=0.
\label{emcl}
\end{eqnarray}

Summing up, the behavior of the scale factor $A(t)$ in the FLRW cosmology
dominated by a perfect fluid given by Eq. (\ref{13}) is determined by
Eqs. (\ref{29}) and (\ref{emcl}). 

The dominant matter-energy contents of the universe in a very early 
phase is expected to be a plasma of quark and gluons (QGP). 
Such a phase would occur when the temperatures exceeded
the value $\Lambda_{QCD} \approx 200MeV$ \cite{mukhanov2005}.  
At this regime the
strong coupling is small and most quark and gluons only interact weakly.
Soft modes could also exist but they would constitute only a small fraction 
of the total energy density. It is not clear by now what would be the
better model to describe the evolution of the spacetime in this phase. 
The production of QGP is expected to occur in 
high energy collisions in the great 
accelerators, as in RHIC and LHC experiments (in BNL and CERN, respectively). 
The results to come from these experiments may shed some light in the
understanding of the very early universe. 
In the following sections we shall examine a working model in which the 
\mbox{Y-M} effective Lagrangian density for quantum chromodynamics
with one parameter background field will be considered to 
describe the matter-energy contents of the isotropic and homogeneous 
FLRW universe. The use of effective Y-M theory in the context of QGP 
and gluon plasma has recently been considered in 
\cite{delorenci2008,baker2009}.

%
%
\section{One-parameter effective Lagrangian for Yang-Mills fields}
\label{IV}
The effective Lagrangian density for quantum chromodynamics (QCD) 
in terms of the parameter background field
$F$ can be presented \cite{savvidy1977,pagels1978,nielsen1979} in the 
form:
\begin{eqnarray}
L_{YM} = -\frac{1}{4}\frac{F}{\bar{g}(\gamma)^2}, 
\qquad \gamma\doteq \log\frac{F}{\mu^4} 
\label{40}
\end{eqnarray}
where the effective coupling $\bar{g}(\gamma)$ is implicitly given by
means of
\begin{equation}
\gamma = \int_{g}^{\bar{g}(\gamma)}\!\!\! dg\; \frac{1}{\beta(g)},
\label{41b}
\end{equation}
with $\beta(g)$ the Callan-Symanzik $\beta$-function,
$\mu$ is the renormalization mass and $g$ the gauge field coupling
constant appearing in the basic QCD Lagrangian. This effective Lagrangian
density should be taken as a classical model that incorporates several
features of the quantum problem. It gives a sufficiently correct 
description of the quantum vacuum, opening a window
to the examination of physically important configurations \cite{pagels1978}.

In fact, there are many invariants of the Yang-Mills fields, 
depending on the specific gauge group \cite{roskies1977}. 
The ansatz used to derive this effective Lagrangian density takes in consideration only the algebraic 
invariant $F$ and imposes consistency with the trace anomaly for the energy-momentum tensor
\cite{collins1977}.

Before analyzing the cosmological implications of this model, some remarks are in order.
The study of the energy density for the effective action associated with
this theory, revels that $E^2>B^2$ would lead to a metastability of the vacuum. 
The interpretation for this behavior can be presented as follows: 
if a region in the system develops a large
$E$ field, it will quickly decay into a configuration where $B^2>E^2$ \cite{pagels1978}. 
Furthermore, we limit our considerations to the case in which only 
the average of the squared magnetic color field $B^2$ survives %
\cite{Dunne,Tajima,Joyce,delorenci2002,baker2009}. 
This is formally equivalent to put $E^2=0$ in Eq. (\ref{2d}). 
With this assumption, the energy conservation law [cf. Eq. (\ref{emcl})]
leads to
\begin{equation}
B(t) = \frac{B_0}{A(t)^2}.
\label{34}
\end{equation}

%

Now, introducing this result in Eq. (\ref{29}) we obtain the following equation
governing the time evolution of $A(t)$:
\begin{equation}
\dot{A}^2 = \frac{\kappa B_0^2}{6}\frac{1}{A^2 \bar{g}^2} - \epsilon.
\label{45}
\end{equation}
The equilibrium solutions of this equation can be obtained by means of
\begin{equation}
\frac{\kappa B_0^2}{6}\frac{1}{A^2 \bar{g}^2} = \epsilon,
\label{46}
\end{equation}
which applies for each possible value of the parameter $\epsilon$ ($0$, $\pm 1$),
corresponding to different topologies of the spacetime.

%
%
\section{Cosmological models}
\label{V}
In terms of an action principle formulation this work deals with
a minimal coupling between gravity and Y-M fields, whose total action 
is given by
\begin{equation}
S =\int {\rm d}^4x \sqrt{-det[g_{\mu\nu}]}(L_R+L_{YM}),
\label{action}
\end{equation}
where $det[g_{\mu\nu}]$ is the determinant of the matrix whose elements are the components of
the spacetime metric, given by Eq. (\ref{metric}), $L_R$ represents the well known 
Einstein-Hilbert Lagrangian density \cite{mukhanov2005}, and 
$L_{YM}$ is the effective Y-M Lagrangian density presented in Eq. (\ref{40}).  

Now, let us examine the solutions for the time evolution of the scale factor
$A(t)$ for the case where the effective coupling  $\bar{g}$ presents the 
form
\begin{equation}
\frac{1}{\bar{g}^2} = 1 +b_0 \gamma. 
\label{47}
\end{equation}
As it will be shown in the next section, this form for the effective coupling correspond to
the first two terms coming from the regime of small coupling with large mean fields 
\cite{savvidy1977,matinyan1978,pagels1978},
where the constant $b_0$ is identified with a beta-function coefficient. The first term, the 
Maxwell-like term, is really negligible at the mentioned regime, but it will be 
maintained here for future reference. 
Now Eq. (\ref{45}) reads
\begin{equation}
\dot{A} = \pm \left(\frac{a}{A^2}\ln\frac{b}{A^2} - \epsilon\right)^{1/2},
\label{52}
\end{equation}
with
\begin{eqnarray}
a &\doteq& \frac{\kappa b_0 B_0^2}{3},
\label{49} 
\\
b &\doteq& \frac{\sqrt{2} B_0}{\mu^2}{\rm e}^{1/b_0}.
\label{50}
\end{eqnarray}
The derivative of Eq. (\ref{52}), which provides
an expression for the acceleration field associated to the function 
$A(t)$, is given by (for $\dot{A}\ne 0$):
\begin{equation}
\ddot{A} = -\frac{a}{A^3}\left(1 + \ln\frac{b}{A^2}\right).
\label{52a}
\end{equation}
It is worth to mention that Eq. (\ref{52}) can be presented as a second order
differential equation in a more suitable form as 
$\ddot{y} + 2a/y + \epsilon = 0$, with $y = A^2(t)$.

Formal solutions of Eq. (\ref{52}) are implicitly given by
\begin{equation}
t - t_i= \pm \int^{A(t)}_{A(t_i)} \left(\frac{a}{z^2}\ln\frac{b}{z^2} -\epsilon \right)^{-1/2} dz,
\label{51}
\end{equation}
which does not contain the equilibrium solutions.
For each value of the parameter $\epsilon$, real solutions can exist only if 
\begin{equation}
\frac{a}{A^2}\ln\frac{b}{A^2} - \epsilon \ge 0.
\label{53}
\end{equation}
In the above condition, the equality gives the equilibrium solutions 
(or equilibrium points) of Eq. (\ref{52}).
In fact, at the equilibrium points we have
\begin{equation}
{\rm e}^{\epsilon A^2/a} = \frac{b}{A^2}.
\label{54}
\end{equation}
As the scale factor $A$ appears squared in this expression, all
solutions will be of the form $\pm A_{\epsilon}$. Nevertheless, the negative solutions
shall not be considered, since they do not correspond to physical results for
the scale factor. 
Summing up, solutions for the time evolution of $A(t)$ are given by Eqs. (\ref{51}) 
and (\ref{54}). 

In what follows the graphic study of 
these solutions will be performed for each given spacetime topology. We are not
assuming any particular normalization for the scale factor. Particularly, $A = 1$
does not imply present time.   

%
\subsection{Euclidean section ($\epsilon = 0$)}
\label{Va}
In the case of an Euclidean section, there is one positive equilibrium solution, which
is given by $A_{0} = b^{1/2}$. It can be seen directly from Eq. (\ref{54}) by setting $\epsilon =0$. 
Now, let us study the behavior of $\dot{A}(t)$ in the neighborhood of $A_0$. From Eq. (\ref{52})
one can see that $\dot{A}(t)$ is real only if $A < b^{1/2}$, otherwise it will
be complex. Furthermore, in the region where $\dot{A}$ is real, it will be positive for
the positive root of Eq. (\ref{52}) and negative for its negative root. In other
words, the equilibrium solution $A_0$ behaves as an attractor for the positive root 
(in the sense that any nearby solution
tend forwards it) and as a source for the negative root (any nearby solution tend away from it). 

\begin{figure}[!hbt]
\leavevmode
\centering
\includegraphics[scale = 1]{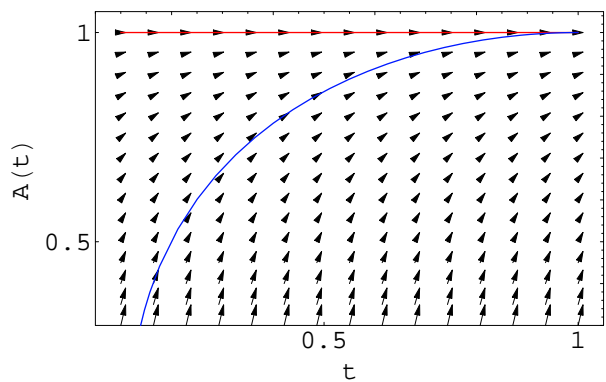}
\vspace{-0.3cm}
\caption{{\small\sf
Slope field for the  positive root of Eq. (\ref{52}) 
in the case of $\epsilon = 0$ (Euclidean section). 
The two solid curves correspond to the equilibrium solution $A_0=b^{1/2}$
(the right line) and a numerical solution of Eq. (\ref{52}) with an initial 
condition satisfying $A(t_i) < A_0$. 
Any solution of the differential equation with initial condition satisfying  
$A(t_i)<b^{1/2}$ evolves towards the equilibrium solution, which behaves as 
an attractor in the plot. There is no solution for which $A > A_0$. The expanding
solution corresponds to a singular and decelerated expanding cosmological model. 
We set $a=1$ and $b=1$. 
}}
\label{slope-solution-num-euclidean-positive}
\end{figure}

The slope field\footnote{The slope field for a differential equation
as $dz/dt = f(t,z)$ consists in the vector field that takes a point
$(t,z)$ to a unit vector with slope given by $f(t,z)$ \cite{hirsch2004}.}  
for the positive root of Eq. (\ref{52}) is presented
in Fig. \ref{slope-solution-num-euclidean-positive}. 
%
%
This figure shows that any initial condition for $A(t)$ satisfying $A(t_i)<A_0$ leads
to a solution presenting a primordial singularity. Singularity is here
understood in the sense that, looking backwards in time, 
$A(t)$ achieves the value zero in a finite interval of time. After this singular origin, 
the scale factor increases in a decelerated rate towards the equilibrium solution 
$A_0$. The two curves plotted in this figure correspond to the equilibrium solution 
(the right line) and a numerical solution of Eq. (\ref{52}) with an initial 
condition satisfying $A(t_i) < A_0$. 

\begin{figure}[!hbt]
\leavevmode
\centering
\includegraphics[scale = 1]{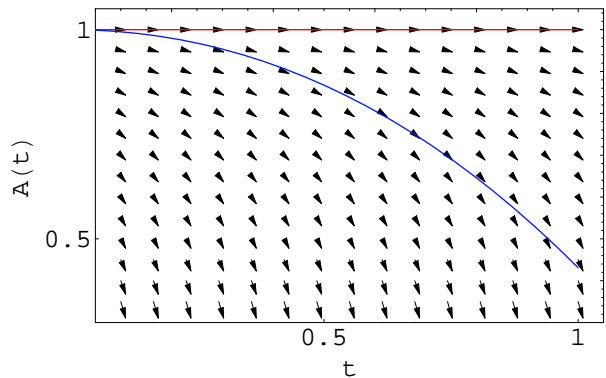}
\vspace{-0.3cm}
\caption{{\small\sf 
Slope field for the negative root 
of Eq. (\ref{52}) in the case of $\epsilon = 0$ (Euclidean section). 
The two solid curves correspond to the equilibrium solution $A_0=b^{1/2}$
(the right line) and a numerical solution of Eq. (\ref{52}) with an initial 
condition satisfying $A(t_i) < A_0$. 
Any solution of the differential equation with initial condition satisfying  $A(t_i)<A_0$ 
decreases quickly to zero. In this case the equilibrium solution 
behaves as a source. The contracting solution corresponds to a singular and 
accelerated collapsing cosmological model. 
We set $a=1$ and $b=1$. 
}}
\label{slope-solution-num-euclidean-negative}
\end{figure}

On the other hand, if the negative
root of Eq. (\ref{52}) is considered, the conclusions are quite different, as shown in
Fig. \ref{slope-solution-num-euclidean-negative}. 
%
%
In this case the
scale factor begins with a finite value, which can not be greater than $A_0$. Then, it
decreases in an accelerated rate to the singularity. This solution corresponds to a collapsing 
singular universe model. In the figure, the equilibrium solution appears as the right
line. The other curve corresponds to a numerical solution  of Eq. (\ref{52}) 
with an initial condition satisfying $A(t_i) < A_0$.       

As the equilibrium solution behaves as a source in the case of the negative root, it is
unstable under small perturbation. Thus, if the system is initially in this 
state, any small fluctuation in the value of $A(t)$ would lead the system to evolve into
a collapsing phase.
In this sense, the two roots could also be considered together. In this case, an initially singular 
universe would expand in a decelerated rate until the
equilibrium solution $A_0$ is achieved. Thus, it could evolve into a singular collapsing phase.

As one can see, the equilibrium point $A_0$ corresponds to an upper limit for the
values the scale factor can take. The value of $A_0$ is associated with the 
parameters appearing in the Lagrangian density and depends on the
specific model for the effective coupling constant. There is no solution presenting
an accelerated expansion in the Euclidean section model.

Since Friedmann equation is invariant under time reversal\footnote{This can 
be understood directly from Eq. (\ref{51}) by taking $t\rightarrow-t$.}, 
solutions for the negative root of Eq. (\ref{51}), as those appearing in 
Fig. \ref{slope-solution-num-euclidean-negative},
can be obtained from the solutions for the positive root simply taking $t \rightarrow -t$ in
Fig. \ref{slope-solution-num-euclidean-positive}. In what follows only 
solutions corresponding to the positive root will be considered in plots.  

It was recently shown \cite{bamba2008} that a non-minimal coupling of gravity with
effective Y-M theory can produce accelerated expansion in the flat model. In fact,
it was claimed earlier, in \cite{zhang1994}, that even for the case of a minimal coupling, 
it would be possible to obtain an accelerated expansion by considering a primordial electric condensate.
Nevertheless, following the reasonings of \cite{pagels1978}, a condensate with only
an electric color field (or even with $E^2>B^2$) would imply in vacuum metastability.
Within the framework considered here, our results confirm that, for a magnetic 
condensate, there is no solution presenting an
inflationary phase in the Euclidean model, if the minimal coupling is assumed.
  
Finally, in the case of Euclidean section, the 
solution presented by Eq. (\ref{51}) can also be presented in terms of the inverse 
error function as
\begin{equation}
A(t) = \sqrt{b}\,{\rm e}^{-\frac{1}{2}{{\rm Erf}}^{-1}\left(C_0 
\pm \frac{2\sqrt{a}}{b\sqrt{\pi}}t\right)^2}
\label{solution-euclidean}
\end{equation} 
where $C_0$ is an integration constant and $\mbox{Erf }(z)$ is the error function 
\cite{gradshteyn1980}
\begin{equation}
\mbox{Erf }(z)=\frac{2}{\sqrt{\pi}}\int _0 ^z {\rm e}^{-x^2} \: dx.
\label{error-function}
\end{equation}
Here we must add again the equilibrium solution $A(t) = b^{1/2}$. The results obtained
for this case (Euclidean section) can also be obtained from the above prescription, i.e.,
by means of Eqs. (\ref{solution-euclidean}) and (\ref{error-function}).

%
\subsection{Closed section ($\epsilon = +1$)}
\label{Vb}
\begin{figure}[!hbt]
\leavevmode
\centering
\includegraphics[scale = 1]{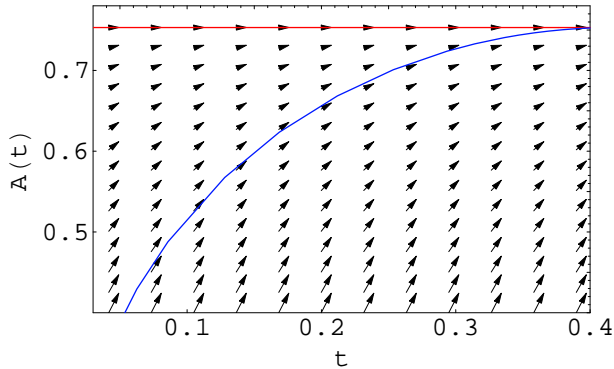}
\vspace{-0.3cm}
\caption{{\small\sf 
Slope field and two solutions for the positive 
root of Eq. (\ref{52}) in the case of $\epsilon = 1$ (closed section). 
The two solid curves correspond to the equilibrium solution $A_{+1}$
(the right line) and a numerical solution of Eq. (\ref{52}) with 
an initial condition satisfying $A(t_0) < A_{+1}$.
Any solution of the differential equation with initial condition satisfying  $A(t_0)<A_{+1}$ 
evolves into the equilibrium solution.
There is no real solution for which $A>A_{+1}$. The expanding solution corresponds to a
singular and decelerated expanding cosmological model. We set $a=1$ and $b=1$.
}}
\label{slope-solution-num-closed-positive}
\end{figure}

In this case the positive equilibrium solution of Eq. (\ref{52}) is given by
\begin{equation}
A_{+1} = \sqrt{a\, W(b/a)},
\label{56}
\end{equation}
where $W(z)$ represents the Lambert function \cite{weisstein1998}. 
Here the behavior of the solutions are quite 
similar to the case examined before, for the Euclidean section. 
The slope field for the positive root of Eq. (\ref{52}) is
presented in Fig. \ref{slope-solution-num-closed-positive}. 
%
%
As occurs in the case of the Euclidean section, the 
equilibrium solution corresponds to an attractor in the
case of the positive root and to a source in the case
of the negative root. In both cases no solutions can be found for which $A > A_{+1}$. 
Figure \ref{slope-solution-num-closed-positive} also presents the
equilibrium solution (the right line) and a numerical solution
corresponding to an expanding model. The latter
is given by Eq. (\ref{51}) with a given initial condition satisfying $A(t_i) < A_{+1}$. 
This solution presents a singular origin, for which $A(t_0) = 0$, evolving into 
a decelerated expansion in the direction of the static configuration determined by $A_{+1}$.

On the other hand, by considering the negative root of Eq. (\ref{52}), 
any given initial condition satisfying $A(t_i) < A_{+1}$ 
leads to an accelerated collapsing model. 

The only difference between the solutions for Euclidean and closed 
sections consists on the value of the equilibrium point. 
Again, there is no solution corresponding to an accelerated expanding model.
For the case of the positive root of Eq. (\ref{52}), any expanding solution
will evolve into a static configuration with $A = A_{+1}$. 
Nevertheless, since the equilibrium solution is unstable for the negative root,
the system in the static configuration could 
evolve into a collapsing phase, provided small fluctuations of the scale 
factor are allowed to occur.

%
\subsection{Open section ($\epsilon = -1$)}
\label{Vc}
\begin{figure}[!hbt]
\leavevmode
\centering
\includegraphics[scale = 1]{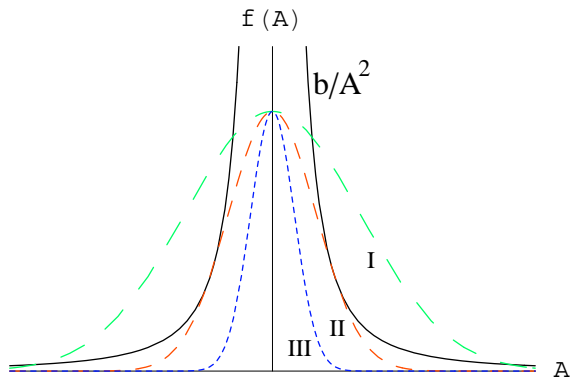}
\vspace{-0.3cm}
\caption{{\small\sf The solid curve corresponds to the plot of the function $f(A)=b/A^2$. The three
dashed curves $I$, $II$ and $III$, correspond to the plots of the function $f(A) = {\rm e}^{-A^2/a}$ 
for $a>{\rm e}b$, $a={\rm e}b$, and $a<{\rm e}b$, respectively. 
The intercepting points between the solid and
the dashed curves occur at the equilibrium solutions, which are given by Eq. (\ref{77}).
}}
\label{fig1}
\end{figure}

In the case of an open section, the equilibrium solutions of 
Eq. (\ref{52}) are given by
\begin{equation}
A_{-1} = i\sqrt{a\, W(-b/a)},
\label{77}
\end{equation}
which is real only if $(b/a) \le (1/{\rm e})$. The mathematical constant ``${\rm e}$'' denotes the
base of the natural logarithm. There are three distinct cases to be
analyzed here, depending on the values of the parameters $a$ and $b$. 
If $a < {\rm e}b$ no equilibrium solutions can be found; if $a = {\rm e}b$ there will
be only one positive equilibrium solution; and finally, if $a > {\rm e}b$ there will
be two positive equilibrium solutions. We remember that these parameters are related with
the physical quantities coming from the Lagrangian densities.
These three branches can be clearly understood with the help of Fig. (\ref{fig1}). In this figure
the solid curve represents the function $b/A^2$ while the dashed curves represent
the function $\exp(-A^2/a)$ for different values of the parameter $a$.  
Equilibrium solutions are determined by the interception between 
the solid and the dashed curves [cf. Eq. (\ref{54})]. 
These curves are denoted by $I$, for $a > {\rm e} b$;
$II$, for $a = {\rm e} b$; and $III$, for $a < {\rm e} b$. 
As one can see, if $a > {\rm e}b$ there will be
two positive and distinct values of $A$ for which these curves intercept. In the limiting
situation where the parameter $a$ attains the exact 
value ${\rm e}b$, there will be only one positive value 
of $A$ for which the curves intercept to each other. 
On the other hand if $a < {\rm e}b$ no coincident point can be found.
%
%

Corresponding to each one of the above mentioned domains, the solutions of the differential
equation (\ref{52}), given by Eq. (\ref{51}), will exhibit different behaviors. 
Therefore, it is worth to examine each case separately.

\subsubsection{$a < {\rm e}b$: no equilibrium solutions}
\label{Vc1}
\begin{figure}[thp]
\leavevmode
\centering
\includegraphics[scale=1]{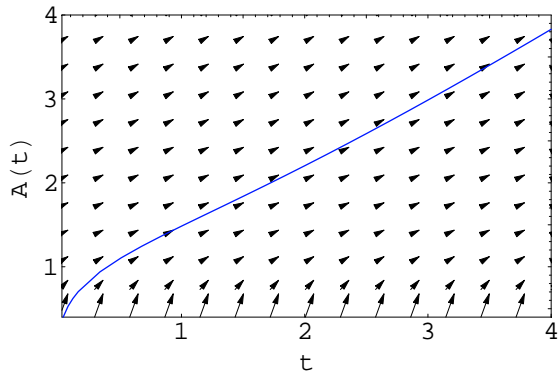}
\vspace{-0.3cm}
\caption{{\small\sf 
Slope field for the positive root of Eq. (\ref{52}) in the case of 
$\epsilon = -1$ (open section) with no equilibrium solutions ($a<{\rm e}b$).
The solid curve represents a numerical solution of Eq. (\ref{52}) with 
a given initial condition and corresponds to an expanding cosmological model with two
distinct phases: a decelerated expansion evolving into 
an accelerated expansion. We set $a=e/2$ and $b=1$.
}
\label{slope-solution-num-open-positive-III}
}
\end{figure}

If $a < {\rm e}b$ Eq. (\ref{77}) does not present any real solution and, as a consequence,
there is no equilibrium point. Taking the positive root of Eq. (\ref{52}) one
can see, from the slope field in Fig. \ref{slope-solution-num-open-positive-III}, that
any given initial condition implies in a singular solution. That
is, for a given instant of time $t_0$ the scale factor attains the singular value 
$A(t_0)=0$. Looking backward, this means that the scale factor achieves
the value zero in a finite interval of time. In this sense $t_0$ can be thought as an
initial time. For $t>t_0$ the scale factor is an increasing function of $t$. 
From Eqs. (\ref{52}) and (\ref{52a}) one can see that this expansion is initially 
decelerated, when $A(t) < ({\rm e} b)^{1/2}$, later evolving into an accelerated
expansion when $A(t) > ({\rm e} b)^{1/2}$.
The transition from the decelerated to the accelerated phase occurs at
a certain transition time $t_T$ given by 
$A(t_T)=(eb)^{1/2}$, which depends on the parameters characterizing
the effective Y-M theory.
The solid curve in this figure corresponds to a numerical solution of Eq. (\ref{52})
and describes the above discussed expanding cosmological model with a primordial 
singularity. 
%

If the negative root of Eq. (\ref{52}) is take into account the opposite behavior is found.
The possible solutions correspond to singular contracting cosmological
models with a decelerated collapsing phase evolving into a singular accelerated collapsing 
phase. 


%
\subsubsection{$a = {\rm e}b$: one equilibrium solution}
\label{Vc2}
\begin{figure}[thp]
\leavevmode
\centering
\includegraphics[scale=1]{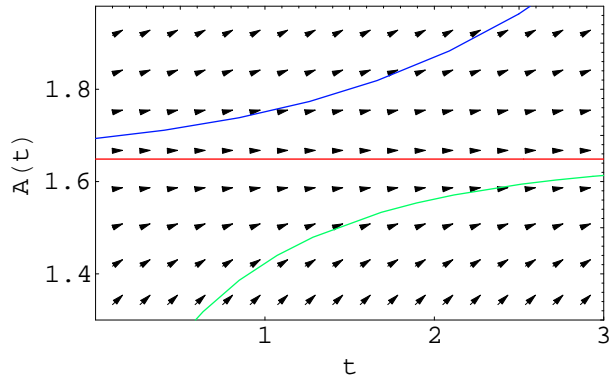}
\vspace{-0.3cm}
\caption{{\small\sf 
Slope field for the positive root of Eq. (\ref{52}) in the case of 
$\epsilon = -1$ (open section) with only one equilibrium solution ($a={\rm e}b$). 
Three numerical solutions of this equation is also plotted. The right 
line represents the equilibrium solution $A_{II}$. Initial conditions such that $A > A_{II}$ lead to 
solutions describing a nonsingular accelerated expanding universe. We set $a=e$ and $b=1$.
}}
\label{slope-solution-num-open-positive-II}
\end{figure}

In this case Eq. (\ref{77}) presents one positive equilibrium solution, which
is here denoted by $A_{II}$. The slope field for the positive root of Eq. (\ref{52})
is presented in Fig. \ref{slope-solution-num-open-positive-II}. Depending on
the initial condition, the system will behave quite differently. 
%
%
For an initial condition satisfying $A(t_i) < A_{II}$, the corresponding solution will be singular
[$A(t_0) = 0$ for an initial time $t_0$], evolving into a decelerated expansion towards
the equilibrium solution $A_{II}$ (the right line in the figure). 
A numerical solution of Eq. (\ref{52}) with a given 
initial condition satisfying $A < A_{II}$ is presented in the bottom of Fig. 
\ref{slope-solution-num-open-positive-II} (the solid curve bellow the equilibrium
solution) and describes a singular cosmological
model presenting a decelerated expansion. 
On the other hand, if $A(t_i) > A_{II}$ the
corresponding solution will be nonsingular [$A(t)>A_{II}\; \forall\; t]$ and 
presenting an accelerated expansion. This cosmological model corresponds
to a nonsingular accelerated expanding universe. 
For a given initial condition this solution is numerically obtained from
Eq. (\ref{52}) and is plotted in Fig. \ref{slope-solution-num-open-positive-II}, 
appearing as the solid curve above the equilibrium solution (the right line).  
These two kinds of solution are 
separated by the equilibrium solution, which behaves as an attractor for solutions 
satisfying $A(t_i) < A_{II}$ and as a source for solutions satisfying $A(t_i) > A_{II}$.
The equilibrium solution itself corresponds to a static cosmological model.
%
As the equilibrium solution is unstable for those solutions 
satisfying the initial condition $A > A_{II}$, a singular decelerated 
expanding phase could evolve into an 
accelerated expanding phase, provided small fluctuations of the scale factor are 
allowed when the system is nearby the equilibrium point at $A=A_{II}$. 

Now, if the negative root of Eq. (\ref{52}) is considered, the possible solutions 
will describe collapsing or static cosmological models. As before, these solutions
can be obtained directly by taking $t\rightarrow -t$ in the above solutions. 

The solutions coming from the positive and negative roots can also be combined. 
For instance a collapsing phase can evolve into an expanding accelerated
phase producing a bounce. Note however 
that this combinations do not correspond to a superposition of solutions. It 
can occur if small fluctuations on the scale factor are allowed to occur
when the system achieves its equilibrium point. 
In this sense, the system initially in the equilibrium state can evolve into an accelerated
expansion if a positive fluctuation of $A(t)$ occurs, or yet, it could evolve into a 
collapsing phase if a negative fluctuation occurs. 

\subsubsection{$a > {\rm e}b$: two equilibrium solution}
\label{Vc3}
Finally, when $a > {\rm e}b$ there will be two positive equilibrium solutions given
by Eq. (\ref{77}), which are denoted by $A_{I}^{a}$ and $A_{I}^{b}$, with
$A_I^a < A_I^b$. We  notice that Eq. (\ref{52}) presents no
real solutions in the interval $A_I^a < A < A_I^b$. Furthermore, 
as the parameter $a$ takes the limiting value ${\rm e}b$ the both equilibrium points 
get the same value $A_{II}$, which corresponds to the case of only one
equilibrium solution, analyzed before.  

\begin{figure}[thp]
\leavevmode
\centering
\hspace*{-0cm}
\includegraphics[scale=1]{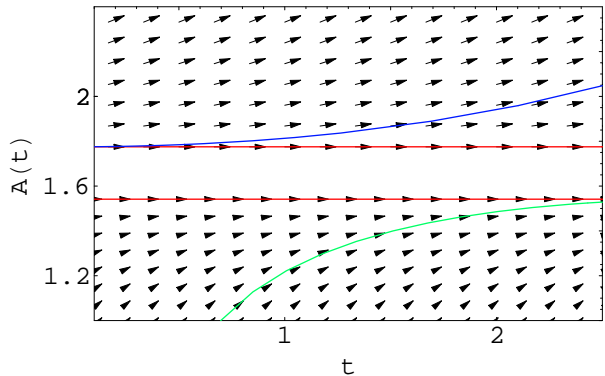}
\vspace{-0.3cm}
\caption{{\small\sf 
Slope field for the positive root of Eq. (\ref{52}) in the case of 
$\epsilon = -1$ (open section). The top corresponds to the 
region where $A(t) \ge A_I^b$, while
the bottom corresponds to the region where $A(t) \le A_I^a$.
There is not real solution between these two limiting values. The equilibrium
solutions $A_I^a$ and $A_I^b$ are represented by the right lines (with $A_I^a < A_I^b$). The two
solid curves correspond to numerical solutions of the positive root of 
Eq. (\ref{52}) for different initial conditions. 
Particularly, any initial condition satisfying $A >A_I^b$ leads to a 
nonsingular accelerated expanding model. We set $a=2e$ and $b=1$.
}}
\label{slope-solution-num-open-positive-I}
\end{figure}

The slope field for the positive root of Eq. (\ref{52}) is presented in 
Fig. \ref{slope-solution-num-open-positive-I}. As one can see, the equilibrium point
$A_I^b$ behaves as a source for those solutions satisfying the initial condition 
$A(t_i) > A_I^b$, while $A_I^a$ behaves as an attractor for those solutions satisfying the
initial condition $A(t_i) < A_I^a$.  
%
\mbox{}
%
For initial conditions satisfying $A(t_i) < A_I^a$, the corresponding solution will be singular
[$A(t_0) = 0$ for a initial time $t_0$] evolving into a decelerated expansion
towards the equilibrium point $A_I^a$.
It corresponds to a singular decelerated expanding cosmological model.
By the other hand, if $A(t_i) > A_I^b$ the
corresponding solution will be nonsingular, with $A(t_0)=A_I^b$, and 
accelerated expanding. It describes a nonsingular accelerated expanding universe. 
Since the equilibrium solution $A_I^b$ is unstable, an initially static phase could evolve
into a expanding one if small fluctuations of the scale factor are allowed.  
These two kinds of solution are 
separated by a region $A_I^a < A < A_I^b$, for which no
solutions can be found.
In Fig. \ref{slope-solution-num-open-positive-I} the equilibrium solutions 
$A_I^a$ and $A_I^b$ are represented by the right lines. Two numerical solutions
of Eq. (\ref{52}) are also presented. They correspond to 
different initial conditions. The curve in the bottom was produced by an initial 
condition with $A(t_i)$ being smaller than $A_I^a$, while the curve in the top
was produced by an initial condition with $A(t_i)$ being greater than $A_I^b$. They
describe, respectively, a singular decelerated expanding model and a nonsingular
accelerated expanding model. 

Again, if the negative root of Eq. (\ref{52}) is considered, the possible solutions 
will describe collapsing or static cosmological models, and they can be obtained 
by taking $t\rightarrow -t$ in the above solutions.

Depending on the equilibrium state the system lies, it can evolve into an 
accelerated expansion [from the positive root of Eq. (\ref{51})] or into an accelerated 
collapsing phase [from the negative root of Eq. (\ref{51})], provided small fluctuations on the
scale factor are allowed.  
In this sense, the solutions coming from the positive and negative roots can also be combined. 
For instance, a decelerated expanding phase can evolve into
an accelerated collapsing phase, or yet, a collapsing phase could evolve into a 
expanding accelerated phase, producing a nonsingular model presenting a bounce.

%
%
\section{Small coupling with large mean-fields}
\label{VI}
For the case of small coupling, the beta function appearing in Eq. (\ref{41b}) can be 
expanded as \cite{gross1973,poli1973,jone1974,casw1974}
\begin{equation}
\beta(g) = -\frac{1}{2}b_0 g^3 + b_1 g^5 + \cdots,
\label{93}
\end{equation}
where $b_0$ and $b_1$ are the usual $\beta$-function coefficients defined in
one- and two-loop orders \cite{pagels1978}.
Now, if we take the limit of large mean fields 
($F/\mu^4 >> 1$) we obtain, from Eq. (\ref{41b}),
\begin{equation}
\frac{1}{\bar{g}(\gamma)^2} = b_0 \gamma - 2\frac{b_1}{b_0} \log\gamma + \cdots
\label{94}
\end{equation}
It has been argued \cite{adler1981} that this expansion may also give the leading
two terms in the effective action for weak mean fields ($F/\mu^4 << 1$), because the
magnitude of the effective coupling in Eq. (\ref{94}) is small in both regions.

Introducing Eq. (\ref{94}) in Eq. (\ref{40}),  we obtain \cite{pagels1978,nielsen1978}
\begin{eqnarray}
L_{YM} \approx - \frac{1}{4}b_0 F \log\frac{F}{\mu^4}.
\label{95}
\end{eqnarray}
The $\beta$-function coefficient $b_0$ is known as the asymptotic 
freedom constant, and can be presented as 
$b_0 = (1/8\pi^2)(11/3)C_2(G)$.
In SU(3) QCD with $N_f$ massless fermion flavors it reduces
to 
\begin{equation}
b_0 = \frac{1}{8\pi^2}\left(11 - \frac{2}{3}N_f \right).
\label{b0-su3}
\end{equation}

As one can see, the effective Lagrangian density describing the regime of
small coupling with large mean fields, given by Eq. (\ref{95}), corresponds
to a particular case of the Lagrangian density considered in the framework discussed
in the last section. In this regime the parameters $a$ and $b$, appearing in
Eq. (\ref{52}), are given by
\begin{eqnarray}
a &=& \frac{\kappa\, b_0\, B_0^2}{3},
\label{69}
\\
b &=& \frac{\sqrt{2}B_0}{\mu^2},
\label{70}
\end{eqnarray}
which are obtained by neglecting the first term in the right-hand-side of 
Eq. (\ref{47}).

It is important to note that Eq. (\ref{95}) does not require further terms in the
approximation as the field becomes larger. As the fields become larger, 
better Eq. (\ref{95}) holds as the dominant contribution to the effective
Lagrangian density.

Before closing this section, some comments on the relationship between the
physical parameters  $b_0$, $\mu$ and $B_0$ and the behavior of the
solutions for $A(t)$ are in order. First, for the cases of Euclidean and closed
sections, the behavior of the solutions are not affected by the values these
parameters can take. On the other hand, for the case of an open section, 
the values of these parameters strongly determine the behavior of
$A(t)$. Particularly, they also make influence on the corresponding model being singular 
or nonsingular. Let us briefly discuss this point in terms of the magnitude of
the primordial magnetic color field for the case of expanding solutions 
in the open section model.   
If $B_0 < (18{\rm e}^2)^{1/2} / (\kappa b_0 \mu^2)$ the corresponding model
will be singular with an earlier decelerated expansion that evolves into an accelerated
expansion phase.
If $B_0 = (18{\rm e}^2)^{1/2} / (\kappa b_0 \mu^2)$ the corresponding
model can be singular or nonsingular, depending on the initial condition $A(t_i)$ be 
smaller or greater than the equilibrium point $({\rm e}\, b)^{1/2} \equiv A_{II}$, 
respectively. In the case of $A(t_i) < A_{II}$ the system will be singular,
evolving into a decelerated expansion until the equilibrium point is achieved. From
this point it can evolve into an accelerated expanding phase or into a collapsing phase (also
accelerated) depending on possible fluctuations of the scale factor $A(t)$. If no fluctuation
is allowed the system will last in the static configuration given by $A_{II}$. On the
other hand, if $A(t_i) > A_{II}$ the corresponding model will be nonsingular and 
accelerated expanding. 
Finally, if $B_0 > (18{\rm e}^2)^{1/2} / (\kappa b_0 \mu^2)$ the expanding solution can be
decelerated or accelerated, depending on the initial conditions for $A(t)$ be smaller than
$A_I^a$ or greater than $A_I^b$, respectively. The singular solution is decelerated 
expanding and can evolve into an accelerated collapsing phase or into a static configuration.
The nonsingular models are accelerated expanding. Such nonsingular models attains its 
minimum $A_{min}$ at the greatest root of 
$\{-(\kappa b_0 B_0^2/3)W[(18)^{1/2}/(\kappa b_0 \mu^2 B_0^2)]\}^{1/2}$. 

%
%
\section{Final remarks}
\label{VII}
WMAP observations \cite{spergel2007,hinshaw2008} have brought several 
implications for cosmology. For instance the parameter related with the spatial 
curvature $\Omega_K$ of the universe is shown to be small, presenting a negative mean 
value.  For some reference values \cite{spergel2007}, it is shown that the combination of 
the present WMAP data with the Hubble Space Telescope data implies in 
$\Omega_K = -0.014 \pm 0.017$; the combination of  WMAP with SNLS 
data implies in $\Omega_K = -0.011 \pm 0.012$; the combination of  WMAP with 
SNGold data implies in $\Omega_K = -0.023 \pm 0.014$. These results seem to 
favor both flat and open models. Further, it was pointed out in \cite{yadav2008}
that the WMAP data contain evidence against the null-hypothesis of 
a primordial Gaussianity. By analyzing the bispectrum of the WMAP up to
the maximum multipole $l_{max} = 750$ they \cite{yadav2008} found 
$27 < f_{NL} < 147$, with $95\%$ of confidence level. This result disfavor
canonical single-field slow-roll inflation predictions of $f_{NL} =0$, and
suggests that alternative models for an early inflationary phase should be 
considered.

In this work the cosmological implications of the minimal coupling between
gravity and effective Yang-Mills theory was examined. An average procedure on 
matter fields was adopted in order to provide a possible way to describe 
an isotropic and homogeneous FLRW cosmology. The behavior of all solutions 
for the time evolution of the scale factor $A(t)$ were graphically examined and the 
main results can be summarized as follows.
Models presenting Euclidean (flat) or closed (positive curvature) spatial section 
can describe a decelerated expanding universe, an accelerated collapsing universe, or
yet a static universe. In all cases, but the static one, these models are singular.
Only models presenting an open spatial section (negative curvature) can support an accelerated
expanding universe. Further, depending on the initial conditions and on the 
values of the parameters $a$ and $b$ that characterizes the system, the corresponding model
can be singular or nonsingular. The expansion will be accelerated if $\ddot{A} > 0$, which occurs
provided $A^2 > {\rm e} b$. 
For the case of $a < {\rm e} b$ the resulting model is singular 
and evolves into a decelerated expansion until $A$ achieves the value 
$({\rm e} b)^{1/2}$ in a finite time. At this point $\ddot{A} = 0$. 
Then, it evolves into an accelerated expanding phase. 
If $a = {\rm e} b$, exactly, there is one equilibrium
point given by $A^2 = {\rm e} b$. Solutions with initial conditions 
$A(t_i) > ({\rm e} b)^{1/2}$ result in a nonsingular and accelerated expanding model, 
while solutions with initial conditions $A(t_i) < ({\rm e} b)^{1/2}$ result in a 
singular model presenting a decelerated expansion.
Finally, if $a > {\rm e} b$ there will be two equilibrium solutions. 
The first one $A_I^a$ smaller than $({\rm e} b)^{1/2}$ and
the second one $A_I^b$ greater than $({\rm e} b)^{1/2}$. Between them, 
there is no physical solution for $A(t)$. 
In this case, for any initial condition satisfying $A(t_i) < A_I^a$, 
the corresponding solution leads to a singular model presenting a decelerated expansion. 
Here the expansion occurs until $A(t)$ achieves the static configuration at $A_I^a$. 
Since this lower equilibrium point is unstable for solutions 
coming from the negative root of Eq. (\ref{52}),
the system can evolve into an accelerated collapsing phase.
Otherwise, if $A(t_i) > A_I^b$ the corresponding 
solution leads to a nonsingular cosmology presenting an accelerated expansion. In this
case $A(t)$ attains its minimum value at $A_I^b$ in a time $t_0$.
Solutions presenting collapsing phases can be obtained from the negative root of Eq. (\ref{51}),
and if combined with the expanding solution, provide a mechanism for a bounce. 
It is worth to stress that, considering the assumed framework, 
solutions presenting accelerated expansion only appear in an open universe model.
Thus, it provides a possible mechanism for an early inflationary phase produced by a vector field. 
The solutions can be singular or nonsingular, depending on the initial
conditions and also on the values that the physical parameters can take. 

Let us pick up a specific application of the results obtained
in this manuscript. Let us take, for instance, a nonsingular expanding solution
from the open-section model with the matter fields described by the Y-M Lagrangian
density in the regime of small coupling with large mean fields [as described in 
Section \ref{Vc3}, with the parameters $a$ and $b$ given by Eqs. (\ref{69}) and (\ref{70})]. 
In this case the universe would come from a previous
decelerated contracting phase, which is a solution coming from the negative root
of Eq. (\ref{52}), and after achieving a minimum volume, determined by $A_I^b$, would 
expand in an accelerated rate. During this expanding phase the strong
energy dominance condition is violated and, mathematically, the accelerated phase would last
forever. Nevertheless, Eq. (\ref{34})
shows that as $A(t)$ increases, $B(t)$ decreases. Thus, the system may achieve a phase in
which the regime of large mean-fields does not apply. This fact shows that
a mechanism for a graceful exit is supported by this framework.
In fact, when $F$ approaches the value $\mu^4$, the dominant term in the effective Lagrangian
would be Maxwell-like. This is formally equivalent to neglect the second term in Eq. (\ref{47}), 
which would lead to a phase of decelerated expansion. In this sense, a
inflationary phase of a nonsingular homogeneous and isotropic universe \footnote{In the singular model, 
an additional phase of decelerated expansion appears before the inflationary phase.} 
with a graceful exit appears naturally in the minimal coupled Yang-Mills 
fields in an open FLRW cosmology. 
Matematically, as consequence of considering Y-M matter field as
the dominant energy contents, all the accelerated phases
described in the open section models last forever. However, as suggested
above, these solutions should be considered only as describing an
early phase of the universe. As the universe expands the plasma
is expected to evolve into another form of matter, which would
dominate a next phase of the evolution. In this
case, a complete scenario should consider the transition between this
primordial phase to the next one, as radiation for example. In the 
case of a singular and accelerated expanding model (as presented
in Section \ref{Vc1}) the amount of inflation is a quantity of interest to
be derived. This quantity can be obtained as $N = \log[A(t_{end})/A(t_{begin})]$
where $t_{end}$ and $t_{begin}$ represent the time coordinate at the end and at
the begin of the inflation phase, respectively. 
Denoting the critical value of the magnetic field by $B_{cr}$, bellow 
which radiation would be the dominant form of matter, we obtain 
$N = (1/2) \log (\mu^2/ \sqrt{2}{\rm e}B_{cr})$. 

Closing, few remarks are in order. 
First, the results
obtained in this manuscript presenting inflationary solutions only occur in models
with open spatial section \footnote{For the case of an electric condensate in a flat model, a 
solution presenting inflation in minimal Y-M theory was proposed in \cite{zhang1994}}. 
Inflationary models obtained from non-minimal Y-M theory have been considered  
only for flat spatial section \cite{bamba2008}. One important feature
of the solutions presenting an inflationary phase in the minimal coupling is that
such phases can also appear in the context of nonsingular models. Thus, in such cases
the amount of inflation is not a fundamental aspect to be considered in the
solution of the initial conditions problems stated by standard cosmology. 
The motivation to work with models with non-flat spatial sections may be 
set by further observations from WMAP and PLANCK projects. It should 
be stressed that the spatial curvature has currently
been found to be small but not necessarily zero, and the consequences of a non-zero
curvature (even really small) can be relevant in the evolution of the universe, 
as shown here. 
Second, all the solutions presenting accelerated phases obtained in the context of miminal
Y-M theory may be naturally obtained as limiting cases from non-minimal couplings 
with an open spatial section. Therefore, if non-flat models is considered in
non-minimal coupling cases our results should be recovered in the corresponding
regime. 
Finally, the anisotropies in CMB found in current data indicates that the early 
phase of the universe was not perfectly smooth. This fact must be connected 
with the inhomogeneities in the distribution of galaxies on large scales, 
as measured by Two Degree Field Galaxy Redshift Survey \cite{colless2001}. 
The very relationship between large scale structure and
CMB is still an exciting issue in cosmology. 
In order to apply the results obtained here in the understanding
of structure formation, the perturbations around the smooth background should be
considered for each particular solution. In this context the power spectrum
in the Fourier space would be the most important statistic. 
In the context of the solutions presented in this manuscript, two different
situations could occur. If singular models are considered, inflation
would be the responsible for the generation of scalar and tensor
perturbations. By the other hand, if nonsingular models are considered,
the primordial perturbations could possibly be produced in a previous contracting
phase, thus evolving into the expanding phase by means of the bounce. 
The derivation of the perturbed Einstein equations in the context
of minimal Y-M theory and its application in the study of the above mentioned 
aspects deserve future investigation.

%
%
\acknowledgments
D-S. Huang, J. Jalilian-Marian and S-Y. Li are acknowledged for
valuable discussions on the Yang-Mills effective theory. 
This work was partially supported by the Brazilian research agencies
CNPq and FAPEMIG.

%
%

\end{document}